\documentclass[apj]{emulateapj}
\usepackage{color}
\usepackage{epsf}
\usepackage{ulem}
\usepackage{epsfig}
\usepackage{graphicx}

\newcommand{\gadget}{\mbox{\sc gadget-2}}
\newcommand{\turtlebeach}{\mbox{\sc turtlebeach}}
\newcommand{\sunrise}{\mbox{\sc sunrise}}

\begin{document}
\title{Molecular Disk Properties in Early-Type Galaxies}
\shorttitle{Molecular Disks in Early-Types}

\author{
X. Xu\altaffilmark{1},
D. Narayanan\altaffilmark{2,3},
C. Walker\altaffilmark{1}}

\begin{abstract}
We study the simulated CO emission from elliptical galaxies formed in
the mergers of gas-rich disk galaxies. The cold gas not consumed in the
merger-driven starburst quickly resettles into a disk-like configuration.
By analyzing a variety of arbitrary merger orbits that produce a range
of fast to slow-rotating remnants, we find that molecular disk formation
is a fairly common consequence of gas-rich galaxy mergers.  Hence, if a
molecular disk is observed in an early-type merger remnant, it is likely
the result of a ``wet merger'' rather than a ``dry merger''. We compare
the physical properties from our simulated disks (e.g. size and mass)
and find reasonably good agreement with recent observations.  Finally, we
discuss the detectability of these disks as an aid to future observations.
\end{abstract}

\keywords{Galaxies: Elliptical and Lenticular, cD---Galaxies: evolution---Galaxies: Kinematics and Dynamics---ISM: molecules}

\altaffiltext{1}{Steward Observatory, University of Arizona,
                933 N. Cherry Ave., Tucson, AZ 85721; xxu@as.arizona.edu}
\altaffiltext{2}{Harvard-Smithsonian Center for Astrophysics, 60 Garden St MS 51, Cambridge, Ma 02138}
\altaffiltext{3}{CfA Fellow}

\section{Introduction}\label{sec:intro}
The ``red and dead'' nature of early-type galaxies such as ellipticals
and lenticulars has long been part of the standard lore. The
general lack of cold molecular gas conducive to the formation of stars
in these galaxies \citep{Lea91} is typically considered the reason for
their redder colours \citep{Bea04}. Recently, however, a number of
studies have shown that some molecular gas can exist in early-types,
oftentimes in a disk-like configuration \citep{Iea96, R97, Wea97, Y02,
  Oea05, Dea05, Y05, Yea08}.

A wide variety of mechanisms have been proposed for the origin of
molecular disks in early-type galaixes.  These include external origins
such as gas accreted from the intergalactic medium \citep{KB05} or gas
remnant after a merger \citep{Lea91, KR96}. Some theories have
additionally speculated an internal origin, i.e. from the mass loss of
evolved stars \citep{FG76, Cea91}. Since star formation occurs in the
presence of molecular gas, the properties of these disks can be used to
study the weak star formation that occurs in as many as 30\% of local
early-type galaxies \citep{Yea05, Kea07}. Their kinematics and dynamics
can also be used to model and infer the potentials of their host galaxies
\citep{dZF89, Cea00}.


Given that ellipticals are thought to form in the merger of two galaxies
where gravitational disruption is prevalent \citep[e.g. ][ and references
therein]{Cea06}, the idea that either stellar or molecular disks may exist
in elliptical galaxies seems counterintuitive.  There is, however, some
theoretical evidence to suggest that stellar disks are born after galaxy
mergers \citep{BH96,Rea06,RB08,Hea09}.  Since stars form from molecular
gas, it is conceivable that molecular disks may also form after ``wet''
(gas-rich) galaxy mergers (as opposed to ``dry'' or gas-poor mergers).

In this paper we use smoothed particle hydrodynamic (SPH) simulations
of gas-rich major mergers combined with molecular CO line radiative
transfer calculations in order to investigate the potential formation
of molecular disks in early-type galaxies.  This approach allows us to
compare synthetic CO observations of our model galaxies, which span
the full range between fast and slow-rotators, with data from the
literature. Although major mergers are rare events in the context
of LCDM \citep{FM08, GW08, Sea08}, the simulations presented here
can still be viewed as useful limiting cases. Our paper is organized
as follows: In \S\ref{sec:mods}, we describe our numerical methods;
In \S\ref{sec:results} and \S\ref{sec:disc} we discuss the formation
of molecular disks in our simulations and compare these with recent
observational results, and in \S\ref{sec:theend} we conclude.

\section{Simulations and Models}\label{sec:mods}
\subsection{Simulations}
We model major mergers between two disk galaxies using a modified version
of the publicly available N-body/SPH code \gadget \ \citep{S05}. Detailed
descriptions of the parent galaxies as well as the multiphase ISM, star
formation and blackhole feedback can be found in \citet{SH02, SH03} and
\citet{ Sea05a}. We refer the reader to these works for a more detailed
description of the simulations. The most relevant aspects to this work
will be summarized briefly below.

The parent disks in our simulations have dark matter halos initialized
with \citet{Hernquist90} profiles. They have a concentration index of
$c=9$, a spin parameter of $\lambda=0.033$ and a circular velocity of
$v_{200}=160$km/s. A total of 120,000 dark matter particles and 80,000
disk particles were used to model each progenitor galaxy. The disk
consists of 40\% gas by mass, with the remainder being collisionless
star particles. The total (halo and baryonic) mass of a parent galaxy
is $\sim2\times 10^{12} M_\odot$. Softening lengths of 100 pc and 200 pc
were used for baryons and dark matter respectively.

The key evolutionary steps during the major merger simulations are
as follows. After the initial passage of the merging galaxies, tidal
torquing drives gas into the central regions \citep{BH91, BH96}, leading
to a period of enhanced star formation rates. As the galaxies undergo
final coalescence during the second passage, the merger-driven starburst
drives the galaxy through a LIRG/ULIRG phase \citep{Yea09}. A combination
of gas consumption and AGN feedback quench the star formation, and the
final merger product evolves passively into a red elliptical galaxy
\citep{Sea05b}.  We analyze the results of the simulations for four
identical mergers which vary only in orbit.  The merger orbit angles
are chosen arbitrarily, and are given in Table \ref{tab:disktab}.

\subsection{Radiative Transfer} \label{sec:radt}
In order to investigate the simulated CO properties of the remnants
of our model mergers, we post-process the SPH output using the 3D
non-LTE molecular line radiative transfer code, \turtlebeach
\ \citep{Nea06,Nea08}.

The molecular gas fraction in galaxies is determined by various different
properties such as metallicity, dust content, interstellar radiation
fields, density and temperature \citep{Hea71, Pea06}. However, since
the spatial resolution is limited in our simulations, this cannot be
calculated explicitly as the locations of individual stars and gas clouds
are unknown. Hence, we assume that half of the cold neutral gas in each
grid cell is atomic and the other half molecular. This is consistent
with studies of nearby star-forming galaxies (e.g. \citet{Kea03}). We
also examined a sub-sample of our simulations by scaling the H$_2$ gas
fraction by the ambient interstellar pressure according to local scaling
relations \citep{BR06}.  In the simulations examined here, the differences
in the two methods for assigning the molecular gas mass are negligible.

The molecular abundance levels are set uniformly at Galactic values
\citep{Lea96}. The relatively high abundance of CO in the galaxy
($\mathrm{CO/H}_2 = 1.5\times 10^{-4}$, \citet{Lea96}) in comparison
to other $\mathrm{H}_2$ tracers such as HCN, CS and $\mathrm{HCO}^+$,
makes it the best observational tracer for molecular hydrogen. Hence,
we focus on the radiative transfer of CO, particularly the readily
observable CO (J=1-0) transition in this work.

\turtlebeach \ calculates the molecular level populations, and hence the
source functions by considering both radiative excitations/deexcitations
as well as collisions with H$_2$.  In practice, a solution grid to the
level populations is guessed and model photons are emitted isotropically
via a Monte Carlo sampling. By assuming statistical equilibrium and
balancing radiative and collisional excitation, de-excitation and
stimulated emission, the level populations are updated until convergence
is reached. The source function can be subsequently determined. The
emergent intensity is then calculated by integrating the equation of
radiative transfer through the model grid. For further details on the
underlying algorithms, please see \citet{Nea08}.

\section{Results}\label{sec:results}
\begin{figure}
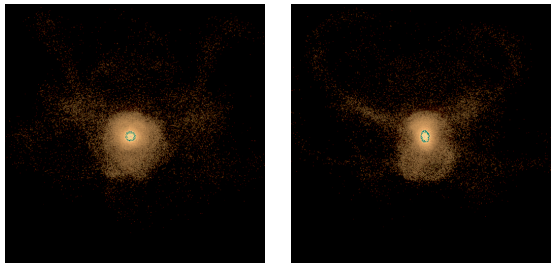

\hspace{1cm}
\centering 
\begin{tabular}{cc}
\epsfig{file=d3e_80_JHK_Cam1.epsi, height=0.4\linewidth, width=0.4\linewidth, clip=}&
\epsfig{file=d3e_80_JHK_Cam3.epsi, height=0.4\linewidth, width=0.4\linewidth, clip=}
\end{tabular}
\caption{Synthetic $JHK$ colours mapped onto $RGB$ for two arbitrary
  viewing angles of our fiducial simulated galaxy, model d3e.  The images
  were produced with the publicly available dust radiative transfer code,
  Sunrise.  The variable input parameters included a stellar birthcloud
  covering fraction of $f_{\rm PDR}=0.3$ and an evolving dust mass with
  the metallicity following a dust-to-metals ratio of 0.4. Evidently,
  the model galaxy has an elliptical morphology.
\label{figure:sunrise}}
\end{figure}

To determine when the product of the major merger has evolved into an
elliptical, we examine the star formation rates (SFRs) and stellar
morphologies of the simulated system.  We pick snapshots well after
the merger-induced starburst, when the stars have relaxed into a
spheroid-type configuration to serve as our model early-type galaxies.
The SFRs are typically of order $\la$0.5 $M_\odot$ yr$^{-1}$.  To confirm
the early-type morphology of our simulated galaxies, we have run dust
radiative transfer calculations in post-processing to determine the
synthetic SEDs\footnote{For this, we employ the publicly available code
\sunrise; a full description of the code can be found in \citet{JGC10}.
For the variable input parameters, we chose a stellar birthcloud
covering fraction of $f_{\rm PDR}=0.3$ and a dust-to-metals ratio of
0.4.}. We plot the simulated $JHK$ colours mapped onto $RGB$ for model
d3e (hereafter our fiducial model\footnote{For reference with previous
papers in the literature, this is additionally the fiducial model in
\citet{Nea08} and \citet{Yea09}.}) at two arbitrary viewing angles
in Figure~\ref{figure:sunrise}. A study by \citet{Cea06} using the same
simulations as this work shows that the early-type galaxies formed span
the full range between fast and slow-rotators. Their rotation parameters
($V_{maj}/\sigma$, where $V_{maj}$ is the major axis rotation speed and
$V_{mag}/\sigma>1$ defines a fast-rotator) range between 0.06 to 1.03
when derived from a model including dissipation.

\begin{figure}
\hspace{1cm}
\begin{center}
\epsfig{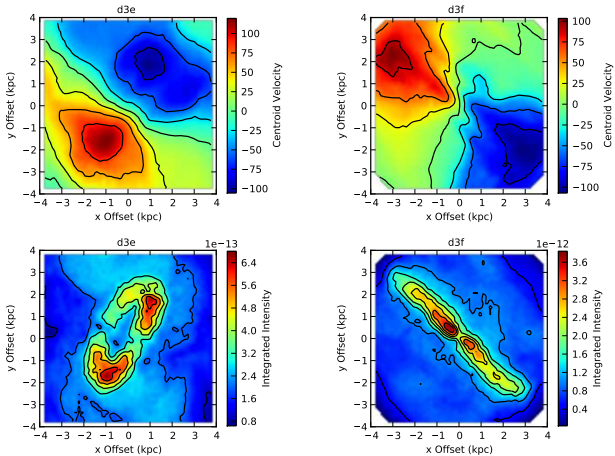}
\caption{(top) Centroid velocity maps in CO (J=1-0) for a nearly edge-on
sight-line through the final elliptical galaxy in our fiducial
model (d3e) and d3f. The units on the velocity are km/s. (bottom)
Same as top, except these are velocity-integrated intensity maps
instead. The units on the velocity-integrated intensity are km/s $\cdot$
erg/s/cm$^2$/Hz/steradian. The clear distinction between the red and
blue velocity peaks in the centroid velocity plot indicates the presence
of a disk.  As we find evidence for molecular disk formation in the
ellipticals produced through all of our merger orbit angles, we conclude
that a variety of merger configurations are conducive to disk formation.
\label{fig:finale}}
\end{center}
\end{figure}

In Figure \ref{fig:finale}, we show centroid velocity maps in CO (J=1-0)
for one sight-line (where the molecular disk is highly inclined) through
our fiducial model and a similarly chosen sight-line through model d3f.
The maps are 8 kpc on a side. The velocity-integrated intensity is shown
for comparison.

We search for evidence of molecular disks in the final elliptical by
examining the CO centroid velocity maps from 100 random viewing angles.
Ordered rotation, as evidenced by clear blue and red centroid velocity
peaks when a disk is highly inclined, is observed along several of the
sight-lines for each merger orbit angle implying that the molecular
gas is in fact located within a rotating disk.  The fact that this is
observed for all of our merger orbit angles indicates that molecular
disk formation may be a relatively common occurence in merger-induced
elliptical galaxy formation.


The possibility of molecular disks being a common occurrence in
elliptical galaxies is interesting.  First, early-type galaxies are
canonicaly thought to be relatively devoid of molecular gas.  Second,
if early-types form in hierarchical merging of dynamically cold disks,
one might naively expect the disks to be disrupted in the merging process.

The molecular disks in our simulations form from gas not consumed in
the merger-driven starburst.  During the merger, gravitational torques
on the gas (by the stars) cause the gas to lose anuglar momentum and
collapse toward the center \citep{BH96,MH96}.  A large increase in
central gas density ensues leading to a burst of star formation. The
remnant gas, originally in the outer regions of the progenitors,
re-virializes into a rotationally-supported disk around the compact
central potential induced by the post-merger stars.  A similar
formation mechanism for post-merger stellar disk formation has been
discussed in detail by \citet{Rea06}, \citet{RB08} and \citet{Hea09}.

\section{Discussion}\label{sec:disc}
\subsection{Comparison with Observations}\label{sec:obs} 
\citet{Yea08}(hereafter Y08) observed the CO (J=1-0) emission in four
early-type galaxies using the BIMA array, which has a primary beam size
of 100''.  Of the four, three showed evidence for molecular disks. The
properties of these three galaxies are listed in Table \ref{tab:ygals}.
We compare our results to their observation by placing our model galaxies
at 17 Mpc ($\sim$ the mean distance of the Y08 galaxies) and convolving
the CO (J=1-0) model spectra with a 100''circular Gaussian beam.
We list the relevant properties derived from our models in the same
table for comparison.


We define the effective radius, $R_e$, to be the radius at which the
stellar mass profile drops from its central value by a factor of 1/$e$.
This implicitly assumes a constant mass-to-light ratio at these smaller
radii \citep{Capea06}.  From comparing values in Table \ref{tab:ygals},
we find the effective radius of NGC 3032 is similar to that of the d3e
merger-remnant. The effective radius of NGC 4459 is in between that of
the d3e and d3k remnants. The effective radius of NGC 4526 is similar
to that of the d3k remnant.

The size of the molecular disks in our simulations is determined in
the following way.  For a near edge-on sight-line for each merger orbit
angle, a line is put down connecting the maximum and minimum velocities
for the molecular disks in the final ellipticals. An interpolation
is performed to find the centroid velocities along this line at
various distances from the center of the disk (taken to be where the
velocity equals zero along the line).  The resulting rotation curve of
the molecular disk is sinusoidal in shape as would be expected for a
rotating disk.  We take the radius corresponding to the extremum of the
rotation curve as the radius of the disk $r_{disk}$. These values are
listed in Table \ref{tab:disktab} and alongside the Y08 values in Table
\ref{tab:ygals}. Reasonable correspondence is seen between the observed
values in Y08 and our models to within a factor of $\sim$2-3.

Finally, the simulated molecular gas masses within the disks are compared
(see Table \ref{tab:disktab}) to those observed.  The study of Y08
uses a CO-to-H$_2$ conversion factor of $3\times 10^{20}$ cm$^{-2}$
(K km/s)$^{-1}$ when deriving their molecular gas masses.  In comparing
these molecular gas masses (see Table~\ref{tab:ygals}), the agreement
seems poorer at first glance.  The typical discrepancies are within a
factor of $\sim$4-10. However, this is about equal to the dispersion
seen in the CO-to-H$_2$ conversion factor in local galaxies \citep[which
ranges by a factor of 2-5 within the Galaxy and by up to an order of
magnitude in other galaxies; ][]{DS98,GS04,Aea96,Bea02}.



The total (baryonic + dark matter) mass of the merger-remnant for each
model is $3.9\times 10^{12} M_\odot$.  The percentage of the total mass
enclosed within the molecular disk radius is given under the column
heading `$f$' in Table \ref{tab:disktab} and are typically of order
$\sim$1-2\%. This result may be applicable in determining the total 
mass of an observed galaxy.


\subsection{Detectability}\label{sec:detect} 
Finally we investigate the detectability of these molecular disks using
two different approaches: from their synthetic CO spectra and from their
centroid velocity maps.

From the 100 arbitrary model viewing angles, we estimate the detectability
of the disk as the fraction of these sight-lines where an obvious
broadened and double-side-peaked line (due to rotation) is visible.
An automated peak finding routine is used to test whether each spectrum
has this characteristic multi-peak profile.  If it does, then we consider
the disk to be detectable.

\begin{figure}
\hspace{1cm}
\begin{center}
\epsfig{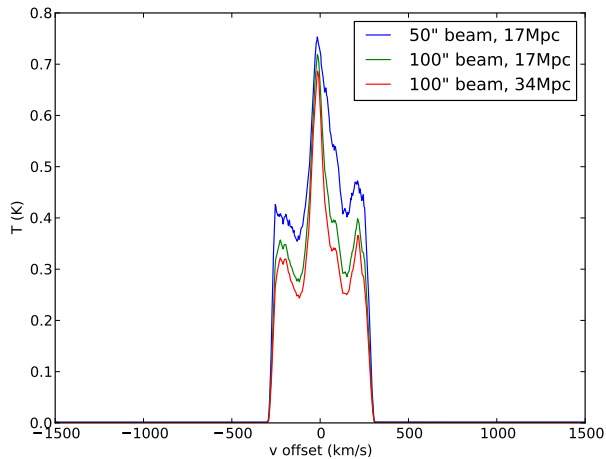}
\caption{CO (J=1-0) spectra for a sight-line through the molecular disk
of model d3m.  The blue line is the expected spectrum of a galaxy 17 Mpc
away convolved with a 50'' beam. The green line is the expected spectrum
of a galaxy 17 Mpc away convolved with a 100'' beam. The red line is
the expected spectrum of a galaxy 34 Mpc away convolved with a 100''
beam. The characteristic signature of a disk is apparent in the broadened
lines with two smaller peaks on opposite sides indicating rotation. We
label these disks as ``detectable'' and define the detectability of the
disk as the fraction of sight-lines through the disk that have these
suggestive profiles. Since the CO line profile may change due to beam
size or distance to the galaxy, we vary these two parameters in the plot
to test the robustness of this characteristic line profile.}
\label{fig:spec} 
\end{center} 
\end{figure}

Examples of such spectra obtained by placing the galaxy at varying
distances and convolving with different beam sizes are shown in Figure
\ref{fig:spec} with 20 km/s smoothing (a typical resolution used by
Y08). The green line in the plot corresponds to the models we compared
to the observations of Y08 in \S\ref{sec:obs} (i.e. distance to galaxy
= 17 Mpc, beam size = 100''). The blue and red lines represent galaxy
distances and beam sizes of 17 Mpc, 50'' and 34 Mpc, 100'' respectively.
Although this is an arbitrary sight-line through the molecular disk of
model d3m, it is representative of all the models in this study.

The detectability of disks using spectral line profiles for the models
comparable to Y08 are given in Table \ref{tab:disktab} under the heading
`$DET_{disk} (\mathrm{spectra})$'. The sightline averaged detectability of
disks (e.g. the fraction of viewing angles over which the characterstic
3-peaked line profile is visible) is $\sim$60-75\%.  This is comparable
to the detection fraction of molecular disks in early-types by Y08
(75\%). The smaller beam size and more distant galaxy cases are detectable
to similar percentage levels (within $\sim2-5\%$).

The spectroscopic CO line profile may change in shape due to beam size
and distance to a galaxy. For example, if the galaxy is isolated and
unresolved in a telescope beam, as the beam size gets larger the line
temperature will decrease while the line shape stays the same. If the
galaxy is resolved, then as the beam gets bigger or smaller, more or
less velocity components within the galaxy are included or excluded
so the line shape may change. Figure \ref{fig:spec} and the similar
detectabilities suggests that this technique for detecting underlying
molecular disks is relatively robust against a variety of modeled galaxy
distances and beam sizes.

Another method that can be used to detect molecular disks is by analyzing
the centroid velocity data. In this case, we can estimate
the detectability of the disk as the fraction of sight-lines in each
merger model that show a single distinct blue velocity peak and a
single distinct red velocity peak, like in Figure~\ref{fig:finale}. If
the centroid velocity map has these distinct peaks, we consider the
disk detectable. Another automated peak finder is used to perform this
search. We find similar detectabilities to the spectra method for the
models comparable to Y08. These are given in Table~\ref{tab:disktab}
under the heading `$DET_{disk} (\mathrm{CV})$' and mostly lie in the
$\sim$60-75\% range as before.

\section{Conclusions}\label{sec:theend}
We study the early-type galaxies formed in \gadget \ simulations of
gas-rich major mergers at four arbitrary merger orbit angles. A 3D
non-LTE radiative transfer code creates 100 CO (J=1-0) spectra for each
merger remnant by ``viewing'' it at arbitrary viewing angles. Molecular
gas disks are detected in all four merger remnants which range between
fast to slow-rotators. This implies that molecular gas disks, thought to
reform after a gas-rich merger through a re-virialization process induced
by the compact central potential of post-merger/post-starburst stars,
may not be a rare occurrence in early-type galaxies. Consequently, if an
early-type merger remnant contains a molecular disk, it is likely that
a wet merger was responsible for its formation rather than a dry merger.

The rare nature of major mergers in the standard LCDM context though,
means that the simulations studied here are likely not perfect analogs
to most early-type galaxies in the local universe. They are, however,
interesting limiting cases in the formation of molecular disks in
early-type galaxies. We would expect that a more typical merger ratio
like 1:3 will leave an S0-like remnant with $<20\%$ of the mass in the bulge
\citep{Hea09}. This implies a smaller disruption to the original gas disks
and hence, a higher probability of having a molecular disk in the remnant.

We compare the molecular disk radii and the molecular gas masses enclosed
within these radii to the observational results of \citet{Yea08}.
Reasonably good agreement between our models and their results are found
to within a factor of $\sim$2-3 for disk radius. The molecular gas masses
differ by a factor of $\sim$4-10. However, these differences are within
the uncertainty of the CO-to-H$_2$ conversion factor.

We find the total mass enclosed within the molecular disk radius to be
$\sim$1-2\% of the total mass (baryons + dark matter) of the galaxy. This
result may be useful for estimating the total mass of an observed galaxy.

We investigate the detectability of underlying molecular disks using
two methods: from CO spectra and centroid velocity maps.  From model CO
spectra, we derive typical disk detection rates of $\sim$60-75\%. This
compares well to the 75\% detection rate of molecular disks in the
galaxies observed by \citet{Yea08}. Using the centroid velocity map
approach, a similar disk detection rate is achieved.

\acknowledgements
The computations in this paper were run on the Odyssey cluster
supported by the FAS Sciences Division Research Computing Group at
Harvard University.



\clearpage

\newcommand{\tableskip}{\\[-8pt]}
\newcommand{\singleline}{\tableskip\hline\tableskip}
\newcommand{\doubleline}{\tableskip\hline\tableskip}
\newlength{\tablespread}\setlength{\tablespread}{30pt}
\newcommand{\dje}{\hspace{\tablespread}}
\tabletypesize{\small}
\def\arraystretch{1.1}

\begin{deluxetable}{ccccccccccc}
\tablewidth{405pt}

\tablecaption{\label{tab:disktab}Simulation orbit angles and derived
properties of product ellipticals}

\tablehead{
\colhead{Model}&
\multicolumn{4}{c}{Orbit Angle}&
\colhead{$r_{disk}$}&
\colhead{$M(r < r_{disk})$}& 
\colhead{$M_{H_2}(r < r_{disk})$}&
\colhead{$f$}&
\colhead{$DET_{disk}$}&
\colhead{$DET_{disk}$}\\
\colhead{}&
\colhead{$\phi_1$}&
\colhead{$\theta_1$}&
\colhead{$\phi_2$}&
\colhead{$\theta_2$}&
\colhead{(kpc)}&
\colhead{($M_\odot$)}&
\colhead{($M_\odot$)}&
\colhead{(\%)}&
\colhead{(spectra)}&
\colhead{(CV)}}
\startdata \doubleline
d3e&30&60&-30&45&2.0&$4.1\times10^{10}$&$5.3\times10^7$&1.1&0.47&0.50\\
d3m&0&0&71&90&2.0&$3.7\times10^{10}$&$2.1\times10^9$&0.95&0.61&0.66\\
d3f&60&60&150&0&3.6&$7.2\times10^{10}$&$3.2\times10^9$&1.9&0.72&0.66\\
d3k&-109&30&71&30&1.8&$3.7\times10^{10}$&$2.3\times10^9$&0.95&0.71&0.75
\enddata

\tablecomments{Models are referred to by the designations under the ``Model''
heading throughout the paper.}
\end{deluxetable}

\begin{deluxetable}{cccccccc}
\tablewidth{405pt}

\tablecaption{\label{tab:ygals}Comparison of properties between Y08 galaxies and our model galaxies}

\tablehead{
\colhead{Property}&
\colhead{NGC 3032}&
\colhead{NGC 4459}&
\colhead{NGC 4526}&
\colhead{d3e}&
\colhead{d3m}&
\colhead{d3f}&
\colhead{d3k}}
\startdata \doubleline
Distance (Mpc)&21.4&15.7&16.4&17&17&17&17\\
$R_e$ (kpc)&0.93&2.7&3.5&1.2&6.0&4.8&3.6\\
$M_{H_2} (r<r_{disk})$ ($M_\odot$)&$5.0\times10^8$&$1.6\times10^8$&$5.7\times10^8$&$5.3\times10^7$&$2.1\times10^9$&$3.2\times10^9$&$2.3\times10^9$\\
$r_{disk}$ (kpc)&1.5&0.67&1.1&2.0&2.0&3.6&1.8
\enddata

\end{deluxetable}


\end{document}